\documentclass[11pt,twoside]{article}
\usepackage{asp2010}

\resetcounters

\begin{document}

\title{Long-term magnetic field monitoring of the Sun-like star $\xi$ Bootis A}
\author{Morgenthaler A.$^1$, Petit P.$^1$, Auri\`ere M.$^1$, Dintrans B.$^1$, Fares R.$^1$, Gastine T.$^{1,2}$, Lanoux J.$^3$, Ligni\`eres F.$^1$, Morin J.$^4$, Ramirez J.$^5$, Saar S.$^6$, Solanki S.K.$^2$, Th\'eado S.$^1$, Van Grootel V.$^{1,7}$}
\affil{$^1$Laboratoire d'Astrophysique de Toulouse-Tarbes, Universit\'e de Toulouse, CNRS, 14 Avenue Edouard Belin, 31400 Toulouse, France}
\affil{$^2$Max-Planck Institut f\"ur Sonnensystemforschung, Max-Planck-Str. 2, 37191 Katlenburg-Lindau, Germany}
\affil{$^3$Centre d'Etude Spatiale des Rayonnements, Universit\'e de Toulouse, CNRS, 9 Avenue du Colonel Roche, 31400 Toulouse, France}
\affil{$^4$Dublin Institute for Advanced Studies, School of Cosmic Physics, 31 Fitzwilliam Place, Dublin 2, Ireland}
\affil{$^5$Instituto de Astronomia, Universidad Nacional Autonoma de Mexico, 04510 Coyoacan, DF, Mexico}
\affil{$^6$Harvard-Smithsonian Center for Astrophysics, 60 Garden St., Cambridge, MA 02138, USA}
\affil{$^7$Institut d'Astrophysique et de G\'eophysique, Universit\'e de Li\`ege, All\'ee du 6 Ao\^ut 17, B 4000 Li\`ege, Belgium}

\begin{abstract}
Phase-resolved observations of the solar-type star $\xi$ Bootis A were obtained using the NARVAL spectropolarimeter at the Telescope Bernard Lyot (Pic du Midi, France) during years 2007, 2008, 2009 and 2010. The data sets enable us to study both the rotational modulation and the long-term evolution of various magnetic and activity tracers. Here, we focus on the large-scale photospheric magnetic field (reconstructed by Zeeman-Doppler Imaging), the Zeeman broadening of the FeI 846.84 nm magnetic line, and the chromospheric CaII H and H$\alpha$ emission.
\end{abstract}

\section{$\xi$ Bootis A (HD 131156A)}
$\xi$ Boo A is the primary component of a visual binary system which follows a 151-yr orbital period \citep{hershey77}. Its mass is $0.86 \pm 0.07 M_{\odot}$ \citep{fernandes98} and its mean effective temperature $T_{eff}$ is known to be 5,550 K \citep{gray94}. \citet*{toner88} determined a rotational period of 6.43 days which we used to determine the rotational phases of our observations. The equatorial velocity projected on the line of sight is equal to $2.9 \pm 0.4$ km/s \citep{gray84}. 

$\xi$ Boo A is a very active star, with irregular fluctuations of chromospheric emission \citep{baliunas95}. Its photospheric magnetic field was first detected by \citet*{robinson80}. A first attempt at modelling the magnetic geometry of the star was proposed by \citet{petit05}, who reconstructed a surface distribution of the magnetic field dominated by a large-scale toroidal magnetic component.

\section{Observations}
The data sets were collected with the NARVAL spectropolarimeter at Telescope Bernard Lyot (Observatoire du Pic du Midi, France). The instrumental setup and reduction procedure is the same as the one described in \citet{petit08}. This instrument has a resolution of 65,000 and covers a wavelength domain from near-ultraviolet (370 nm) to near-infrared (1,000 nm). It provides simultaneous recordings of intensity (Stokes I) and circularly polarized (Stokes V) spectra. One set of spectra was obtained for each year between 2007 and 2010.  The first set was recorded from July 26 to August 10 in 2007, and contains 9 spectra. In 2008, 19 spectra were collected between January 18 and February 15. The third set contains 13 spectra from May 28 to July 5, 2009. The last 9 spectra were obtained between 2009 December 14 and 2010 February 14. For each year, the maximum $S/N$ ratio (around 700 nm) averaged over the time-series is successively 711, 582, 455 and 678.

\begin{figure}[!ht]
\plotone[scale=0.4]{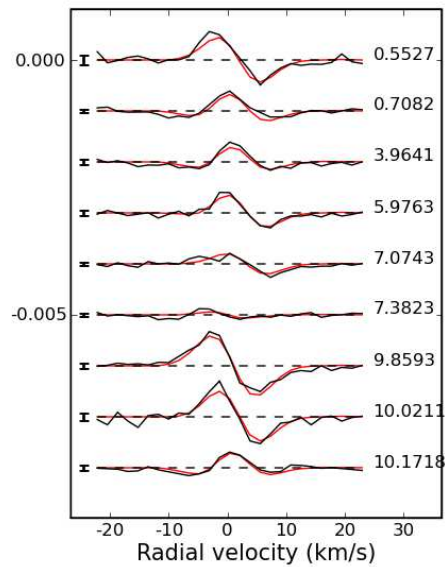}
\caption{2010 Stokes V LSD profiles of $\xi$ Bootis A, after correction of the mean radial velocity of the star. Black lines represent the data and red lines correspond to the synthetic profiles of our magnetic model. Successive profiles are shifted vertically for display clarity. Rotational phases of observations are indicated in the right part of the plot and error bars are illustrated on the left of each profile.}
\label{stokesV}
\end{figure}

\section{Magnetic field maps}
Using a line-list containing about 6,200 lines and matching a stellar photospheric model for the spectral type of $\xi$ Bootis A (G8V), we calculate from the observed spectrum a single, averaged photospheric line profile using the LSD multi-line technique \citep{donati97, kochukhov10}. Thanks to this cross-correlation method, the noise level of the mean Stokes V line profile (Fig.\ref{stokesV}) is reduced by a factor of about 30 with respect to the initial spectrum. The resulting noise levels are in the range $3.0\, 10^{-5}$--$1.6\, 10^{-4} I_c$ (where $I_{c}$ denotes the continuum intensity).

\begin{figure}[!ht]
\plottwo{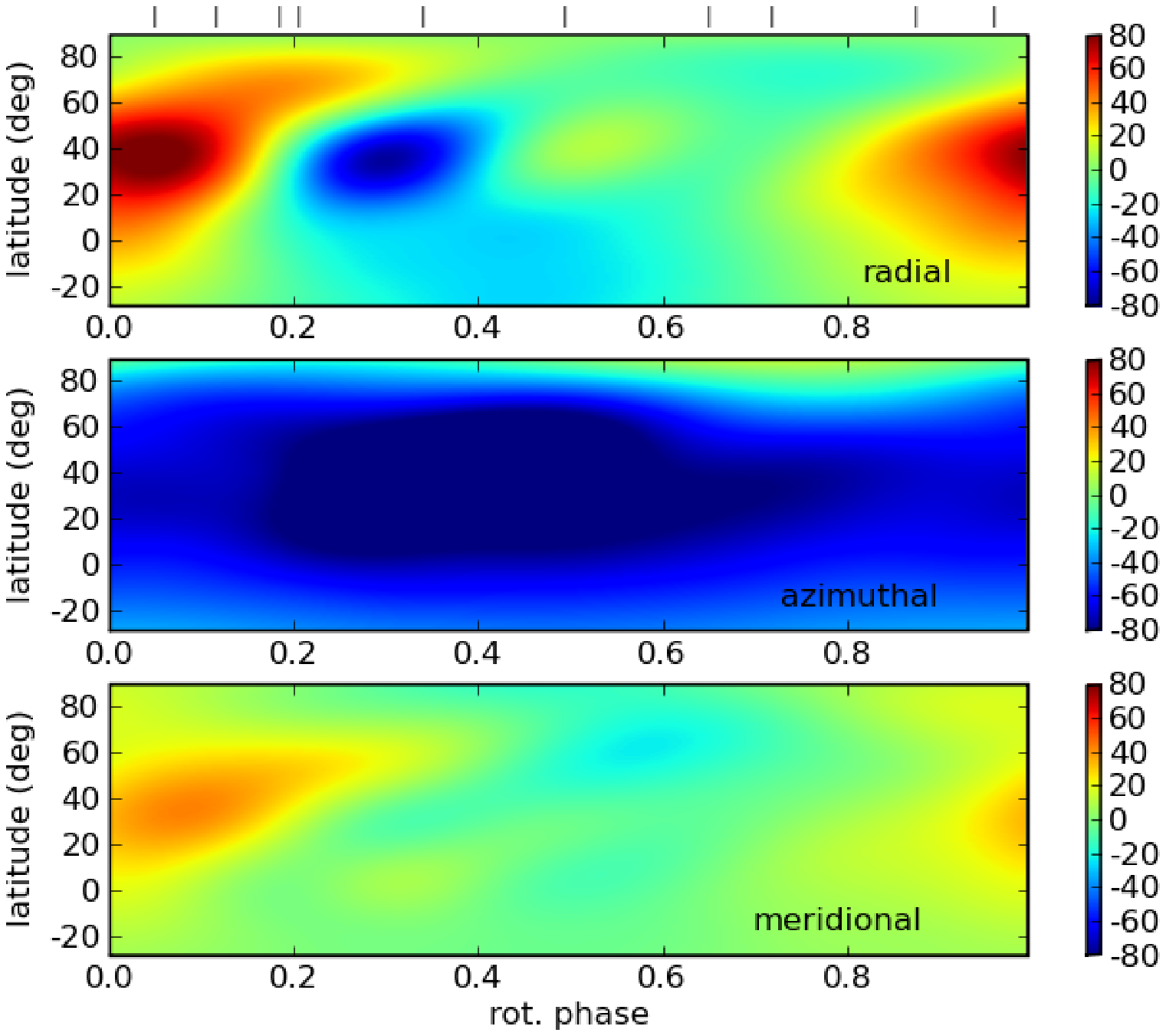}{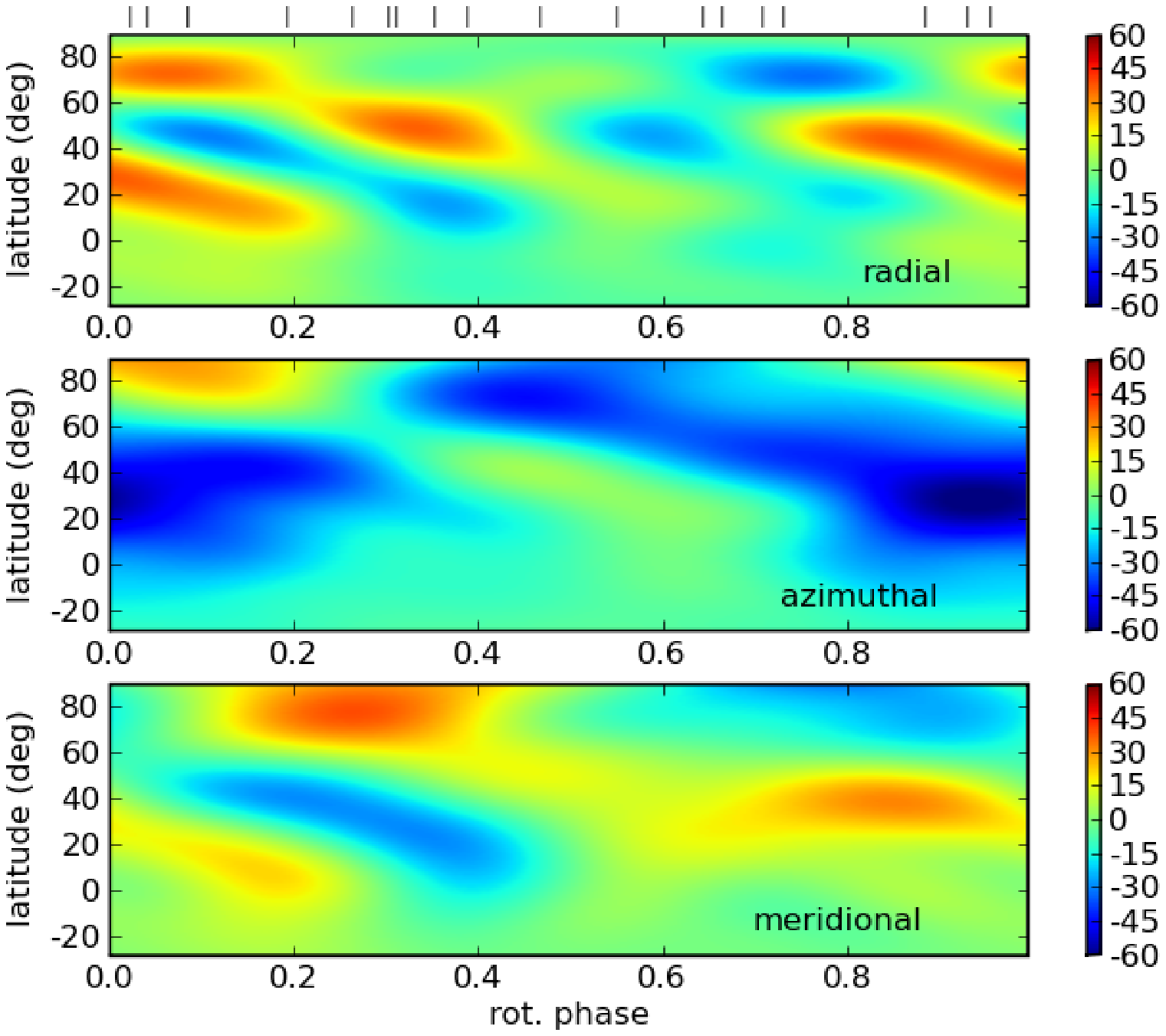}
\plottwo{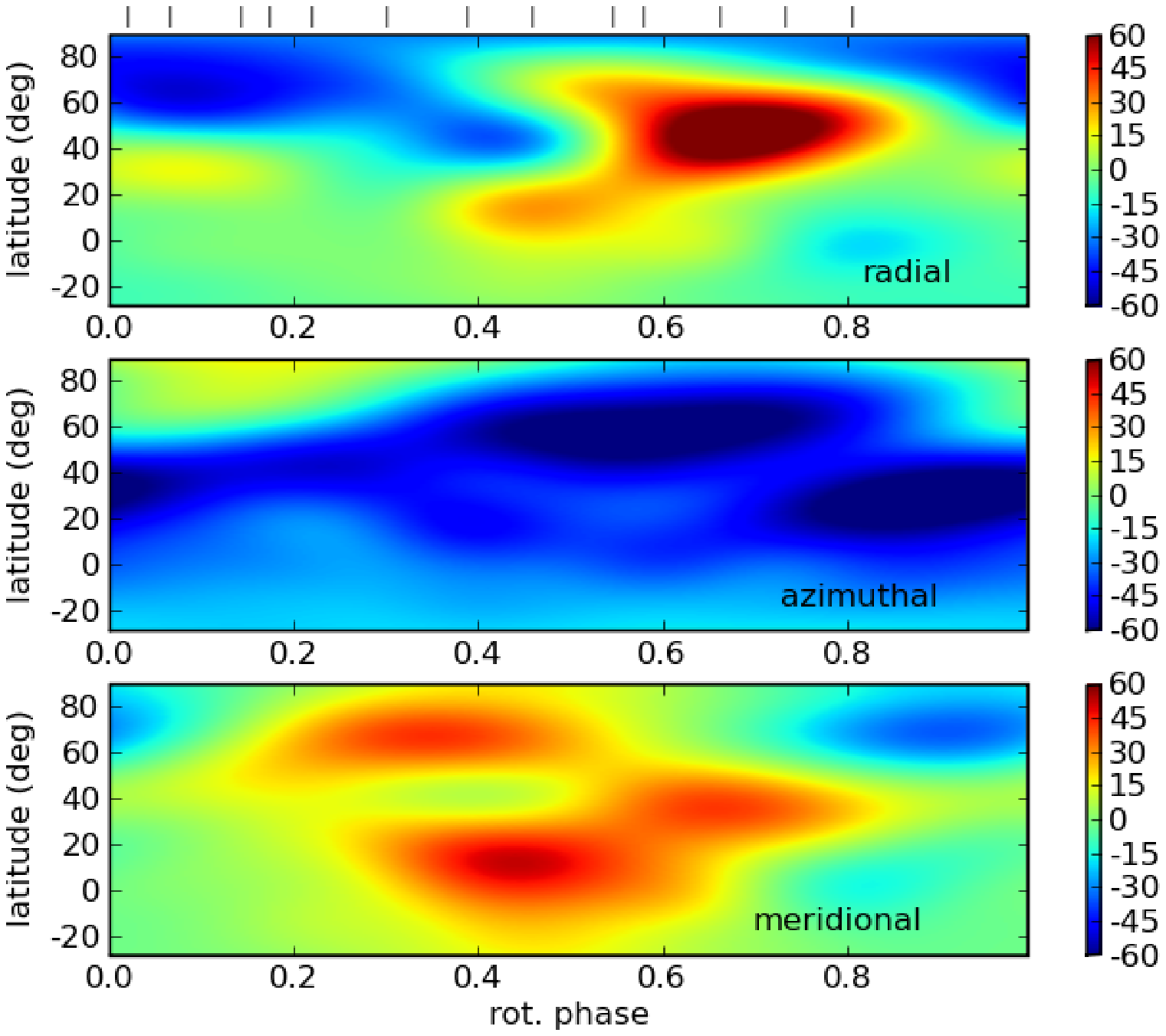}{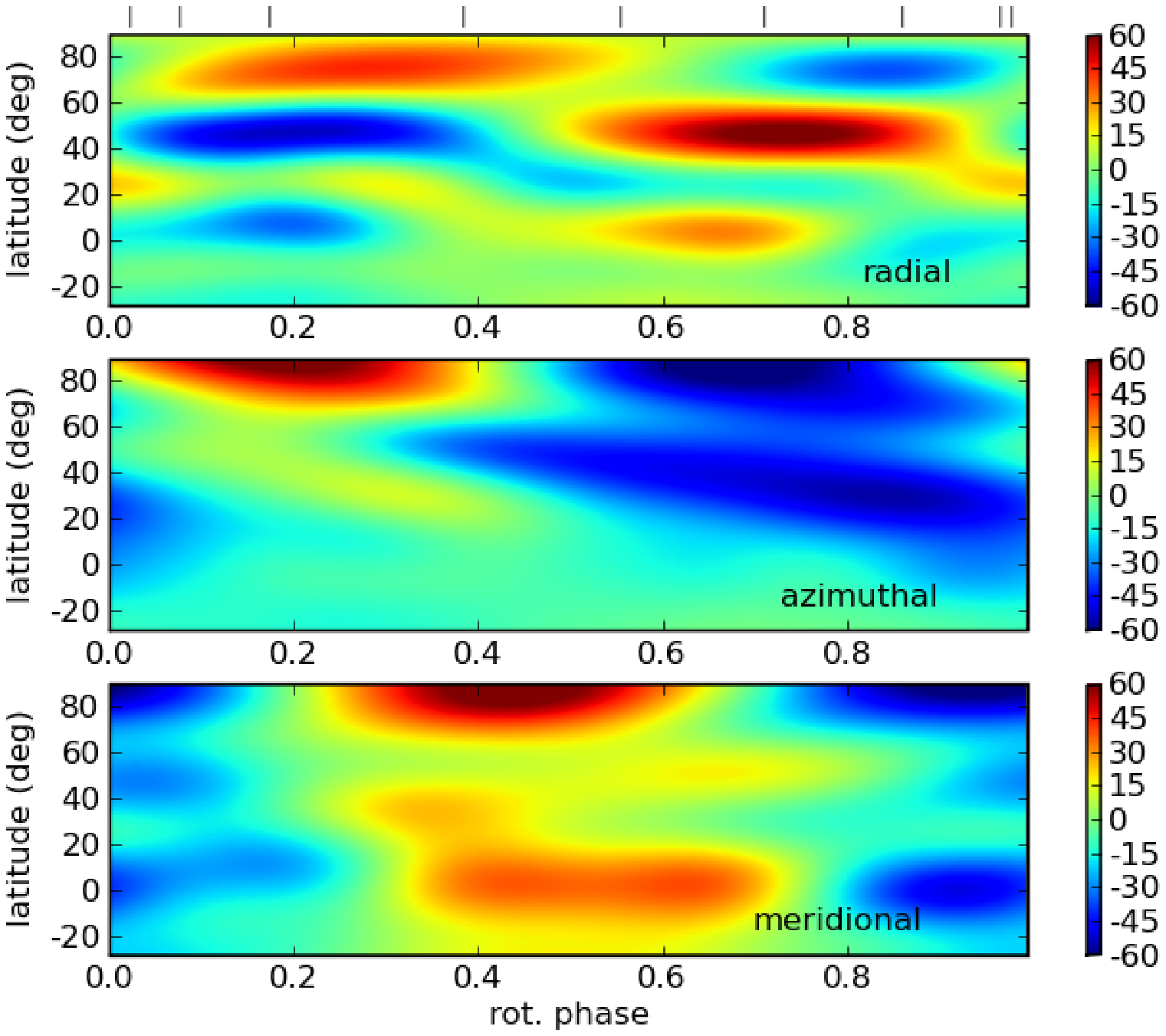}
\caption{Large-scale magnetic field maps of $\xi$ Bootis A for 2007 (top left), 2008 (top right), 2009 (bottom left) and 2010 (bottom right). For each year, the three components of the field vector, in spherical coordinates, are represented from top to bottom (radial, azimuthal, meridional). The magnetic scale is expressed in Gauss (note that the scale is not the same for all years) and the rotational phases of observation are indicated as vertical ticks above each epoch. We take the same phase zero as \citet{petit05}, ie Julian date 2 452 817.41.}
\label{maps}
\end{figure}

Assuming that the observed temporal variability of Stokes V profiles is controlled by the stellar rotation, we reconstruct the magnetic geometry of the star by means of Zeeman-Doppler Imaging (ZDI). We employ here the modelling approach of \citet*{donatibrown97}, including also the spherical harmonics expansion of the surface magnetic field implemented by \citet{donati06} in order to easily distinguish between the poloidal and toroidal components of the reconstructed magnetic field distribution. The obtained maps are displayed in Fig.\ref{maps}.

The global mean field is around 82 Gauss in 2007, 32 in 2008, 53 in 2009 and 38 in 2010. The fraction of poloidal field (wrt toroidal field) is 17\% in 2007, 38\% in 2008, 27\% in 2009 and 32\% in 2010.

\section{Chromospheric activity (CaII H and H$\alpha$)}
\begin{figure}[!ht]
\plotone[scale=0.5]{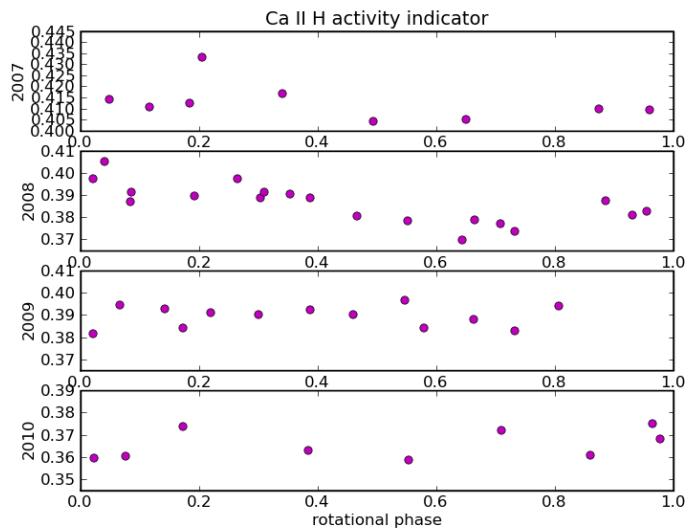}
\caption{CaII H activity indicator as a function of the rotational phase for 2007, 2008, 2009 and 2010 (from top to bottom).}
\label{Nepoque}
\end{figure}

To study the evolution of the chromospheric CaII H emission during a rotation period, we calculate an emission index from our sets of Stokes I spectra with the method described in \citet{wright04} (Fig.\ref{Nepoque}). The index globally decreased between 2007 and 2010 (the mean values for the four years are respectively 0.413, 0.387, 0.390 and 0.366). A rotational modulation is clearly observed in 2008 and 2010. We also reconstruct a similar index to monitor the evolution of the chromospheric H$\alpha$ emission using the same passbands as \citet{gizis02}. A very good correlation (with a Pearson coefficient of about 0.85) shows up between the H$\alpha$ and the CaII H indices.

\section{Zeeman broadening}
We perform a rough estimate of the Zeeman broadening, and of its temporal fluctuations, by monitoring the width of a near infrared FeI line with a high Land\' e factor ($\lambda$ = 846.84 nm, g = 2.493). The line width is calculated for each intensity spectrum. 

\begin{figure}[!ht]
\plotone[scale=0.5]{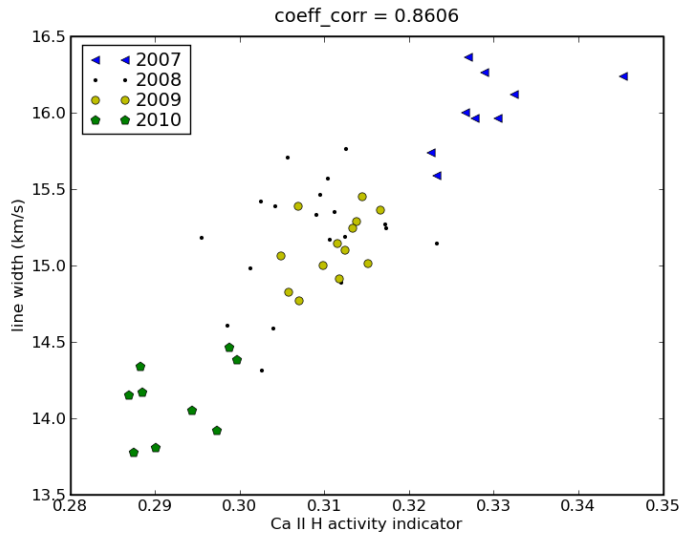}
\caption{Correlation between the widths of the FeI 846.84 nm magnetic line and the values of the CaII H index.}
\label{correlation}
\end{figure}

We observe a decrease between each successive year, in agreement with the similar evolution observed in the chromospheric emission. Moreover, there is a clear correlation (with a Pearson coefficient of 0.86) between the CaII H activity indicator and the widths of the magnetic line, for the whole data set (Fig.\ref{correlation}). We note that such correlation does not exist between the chromospheric index and the width of a neighbouring photospheric line with a low Land\'e factor. We note also that the significant change in Stokes I line width (Fig.\ref{correlation}) implies a significant decrease in the total unsigned magnetic flux between 2007 and 2010.

\section{Discussion}
An obvious decrease of the global magnetic field is visible between 2007 and 2010 in the magnetic maps. The same kind of evolution occurs for the CaII H and H$\alpha$ index during the same period, as well as for the width of FeI 846.84 nm magnetic line. During the same time interval, the fraction of large-scale magnetic energy stored in the poloidal field component is multiplied by a factor of two, while the opposite trend is observed in the unsigned magnetic flux.

During the timespan of our monitoring, the large-scale magnetic geometry of $\xi$ Bootis A does not experience dramatical changes like the global magnetic polarity reversals recently reported for HD190771 \citep{petit09} or $\tau$ Bootis \citep{fares09}. Future monitoring of the star will tell us wether its magnetic evolution is associated to some kind of cyclicity, and if its global magnetic field can undergo polarity switches, as only observed on more massive solar-type stars up to now.

\bibliography{morgenthaler_a}

\end{document}